\documentstyle[aps,prl,twocolumn,epsfig]{revtex}
\newcommand{\beq}{\begin{equation}}
\newcommand{\eeq}{\end{equation}}
\newcommand{\bea}{\begin{eqnarray}}
\newcommand{\eea}{\end{eqnarray}}
\newcommand{\beas}{\begin{eqnarray*}}
\newcommand{\eeas}{\end{eqnarray*}}

\newcommand{\Jpar}{J_{\|}}
\newcommand{\Jperp}{J_{\bot}}

\newcommand{\pdag}{{\phantom{\dagger}}}

\begin{document}
\draft
\wideabs{
\title{Flow equation analysis of the anisotropic Kondo model}
\author{W.~Hofstetter$^{(1)}$ and S.~Kehrein$^{(2)}$\cite{leave}}
\address{$^{(1)}$Theoretische Physik III, Elektronische Korrelationen und
Magnetismus, Universit\"at Augsburg, 86135 Augsburg, Germany
$^{(2)}$Lyman Laboratory of Physics, Harvard University,
Cambridge, MA~02138}
\date{\today}
\maketitle
\begin{abstract}
We use the new method of infinitesimal unitary transformations 
to calculate zero temperature correlation functions in the
strong--coupling phase of the anisotropic Kondo model. We find
the dynamics on all energy scales including the crossover
behaviour from weak to strong coupling. The integrable structure
of the Hamiltonian is not used in our approach. Our 
method should also be useful in other strong--coupling models
since few other analytical methods allow the evaluation of their
correlation functions on all energy scales.
\end{abstract}

}
\section{Introduction}
The Kondo model, originally introduced by Zener \cite{Zener 51}, 
is the most basic Hamiltonian describing the physics of dilute 
magnetic impurities in metals. 
Despite its apparent simplicity, it has become a paradigm 
for complex many--body effects, due to its non--trivial 
strong--coupling behaviour at low temperatures. 
Unprecedented theoretical activity during the 
last four decades was triggered by the seminal work 
of Kondo \cite{Kondo 64}, who showed that perturbation 
theory for an antiferromagnetic exchange coupling diverges at 
low temperatures. Within perturbative scaling 
\cite{Anderson 69,Anderson 70a,Anderson 70b} the
coupling constant grows continuously and eventually diverges. 
This indicates that the impurity spin is screened in the 
ground state. However, the diverging coupling constant also 
implies the breakdown of the perturbative scaling approach in 
the strong--coupling phase. Therefore perturbative scaling does not
provide a systematic and controlled expansion describing the 
weak to strong--coupling crossover. This is a typical situation
in many \emph{strong--coupling} problems.
The development of new, nonperturbative 
methods like Wilson's numerical renormalization group (NRG) \cite{Wilson 75} 
and the Bethe ansatz \cite{Tsvelick 83} eventually led to a 
quantitative understanding of the strong--coupling regime in the Kondo
model. Despite these successes, however, a simple \emph{analytical} 
description of the crossover from a free spin to the Fermi 
liquid regime in an RG type framework was still missing.

In the present work, we will attempt to fill this gap by applying Wegner's 
\emph{flow equation method} \cite{Wegner 94},  
which was recently employed successfully to 
diagonalize a related strong--coupling problem, the quantum sine-Gordon
model \cite{Kehrein 99}. We will mainly focus
on the calculation of equilibrium dynamical 
correlation functions at zero temperature as this provides one of the most
interesting new perspectives for this new approach. The calculation
of correlation functions throughout the crossover region
is notoriously difficult with exact 
methods building on the integrability of the model. There is considerable
interest in theoretical tools that allow their determination in 
strong--coupling models, which provides
the main motivation for our work \cite{Lesage 96}. 
Also the integrability of the
Kondo model is not used in our method and nonintegrable
perturbations can be studied as well.   

We consider the \emph{anisotropic} Kondo Hamiltonian
\bea \label{Hamiltonian_1}
H &=& \sum_k \epsilon^\pdag_k c^\dagger_{k\alpha} c^\pdag_{k\alpha} + 
\Jpar\, c^\dagger_{0\alpha}\, \sigma^z_{\alpha\beta}\, c^\pdag_{0\beta}\, S^z \\
&&+\frac{\Jperp}{2}\, \left(c^\dagger_{0\alpha}\, \sigma^+_{\alpha\beta}\, 
c^\pdag_{0\beta}\, S^- + 
h.c. \right) \ . \nonumber
\eea
Here $\vec S$ is a spin-1/2 degree of freedom and 
$c^{(\dagger)}_{0\alpha} = \sum_k c^{(\dagger)}_{k} / \sqrt{L}$
are creation and annihilation operators for the 
localized electron state at the spin site. Notice that the anisotropic 
Kondo Hamiltonian is equivalent to the dissipative two-state system 
(spin boson model) \cite{Guinea 85,Leggett 87}. 
For the noninteracting conduction band we assume a linear dispersion.
We will treat the model in its bosonized form \cite{Schotte 70}: 
One introduces the bosonic spin density modes
$\sigma(p) = \frac{1}{\sqrt{2|p|}} \sum_{q} \left(c^\dagger_{p+q \uparrow} 
c^\pdag_{q \uparrow} - c^\dagger_{p+q \downarrow} c^\pdag_{q
\downarrow} \right)$ with the commutator
$[\sigma(-q),\sigma(q')]=\delta_{qq'}L/2\pi$ for $q,q'>0$. Here
$L$ is the system size and all other nonvanishing commutators
can be derived using $\sigma^\dagger(q)=\sigma(-q)$. Notice
$\sigma(-q)|\Omega>=0$ on the ground state for $q>0$.
The charge density modes in (\ref{Hamiltonian_1}) decouple 
completely and one only has to look at the spin density part
of (\ref{Hamiltonian_1}) \cite{Kleinfactors}
\begin{displaymath}
H = H_0  - 
\frac{\Jpar}{\sqrt{2} 2 \pi} \partial_x \Phi(0)\, S^z 
 + \frac{\Jperp}{4\pi a}\, \left(e^{i \sqrt{2} \Phi(0)}\, 
S^- + h.c. \right)  .
\end{displaymath}
Here $H_0=\frac{2\pi}{L}\, v_F \sum_{q>0}q\, \sigma(q)\, \sigma(-q)$.
In the following we set $v_F = 1$.
The bosonic field is defined as \cite{Delft 98}
\begin{displaymath}
\Phi(x) = -\frac{2 \pi i}{ L} \sum_{q \ne 0} \frac{\sqrt{|q|}}{ q}\:
e^{-i q x - a|q|/2}\: \sigma(q) \ .
\end{displaymath}
The parameter~$a>0$ generates the UV--regularization of our model.
We are interested in its universal properties at energies
$|E|\ll a^{-1}$.
%
%
The longitudinal spin coupling can be
eliminated by a unitary transformation 
$U=\exp\left(i \mu S^z \Phi(0)\right)$ with $\mu=\Jpar/\sqrt{2}2\pi$.
We arrive at 
\beq   \label{Hamiltonian_compact}
H = H_0 + g_0\, \left(V(\lambda_0, 0)\, S^- + 
V(-\lambda_0, 0)\, S^+ \right)
\eeq
with the coupling constant $g_0 = \Jperp/4 \pi a $
and $\lambda_0 = \sqrt{2} - \Jpar / \sqrt{2} 2 \pi $.
Here \emph{vertex operators} have been introduced
\begin{displaymath}
V(\lambda,x)\equiv \exp\left(i \lambda \Phi(x)\right) \ ,
\end{displaymath}
which are complicated many-body objects for a general value
of the \emph{scaling dimension}~$\lambda$~\cite{Delft 98}. 
Notice that our $V(\lambda,x)$ are 
not normal--ordered. We will also need the Fourier-transformed
normalized vertex operators $C^\dagger_p = \alpha_p^{-1}\,L^{-1/2}
\int dx\,e^{ipx}\,V(\lambda,x)$ with 
$\alpha_p^2 =2 \pi a |p a|^{\lambda^2 - 1} / \Gamma(\lambda^2)$.
Notice $C^\dagger_{-p}|\Omega\rangle=C^\pdag_p|\Omega\rangle=0$
for $p>0$, and 
$\langle\Omega| C^\pdag_p\,C^\dagger_{p'}|\Omega\rangle=\delta_{pp'}\theta(p)$,
$\langle\Omega| C^\dagger_p\,C^\pdag_{p'}|\Omega\rangle=\delta_{pp'}\theta(-p)$
for $|ap|\ll 1$.

Eq.~(\ref{Hamiltonian_compact}) with general initial couplings
$\lambda_0$ and $g_0$ is our starting point for
the flow equation approach. The universal 
low--energy properties of (\ref{Hamiltonian_compact}) are equivalent
to those of the spin--density part of (\ref{Hamiltonian_1}). As
proposed by Wegner \cite{Wegner 94}, we apply
a sequence of infinitesimal unitary transformations to (\ref{Hamiltonian_compact})
in order to make it successively more diagonal. One can set this up
in a differential formulation 
\beq  \label{floweq}
\partial_B H(B) = \left[\eta(B),H(B)\right]
\eeq
with an anti--Hermitian generator $\eta(B)$. $H(B=0)$ is our initial
Hamiltonian (\ref{Hamiltonian_compact}) and all $H(B)$ as obtained by the
solution of (\ref{floweq}) are unitarily equivalent to it. In general
we will have to truncate our system of equations, so this equivalence
will only be approximate. In order to generate a stable sequence
of approximations, we choose $\eta(B)$ such that first (for small~$B$)
offdiagonal matrix elements are removed which couple states with
large energy differences (of order~$B^{-1/2}$), and later more degenerate
states. This is reminiscent of the energy scale separation underlying
perturbative scaling. Similar ideas have independently been developed
by G{\l}azek and Wilson (\emph{similarity renormalization scheme})
\cite{GlazekWilson 93} that contain Wegner's method as a special case.

As the flow parameter $B$ grows, the effective spin flip interaction 
in the interaction part $H_{\rm int}(B)$ will become increasingly nonlocal
and the scaling dimension of the vertex operator will begin to flow.
We take this into account by writing
\beq
H(B)=H_0+\int dx\:g(B,x)\, \left(V(\lambda(B), x)\, S^- + 
{\rm h.c.} \right)
\label{HB}
\eeq
with 
$g(B,x) = L^{-1/2} \sum_p g_p(B)\, e^{i p x}$. Wegner has suggested the
generic choice $\eta^{(1)} = \left[H_0, H_{\rm int}\right]$ for the
generator, as this always removes the off--diagonal matrix elements
in the above energy--scale separated way \cite{Wegner 94}.
Our generator has the following more general form
\bea
\eta &=& \sum_p \eta^{(1)}_p \, \left(C^\dagger_p\, S^- - 
C^\pdag_p\, S^+ \right) \nonumber \\
&& + \sum_{p,q} \eta^{(2)}_{pq}\, \left(C^\dagger_p\, C^\pdag_q  - 
C^\pdag_q\, C^\dagger_p\right) \ . \nonumber
\eea
Here $\eta^{(1)}_p = p \alpha_p g_p$ follows from Wegner's canonical
choice. The second generator part has been introduced since it will
later allow us to generate transformed Hamiltonians with a particularly 
simple structure. The precise form of $\eta^{(2)}_{pq}$ will be 
discussed below.

\begin{figure}[t]
\begin{center}
\epsfig{file=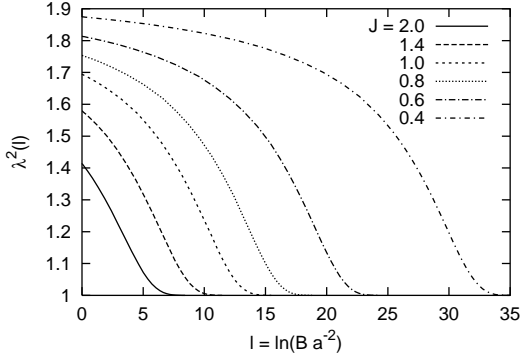,width=0.8\linewidth}
\end{center}
\caption{\label{fig:lambda2}Flow of the scaling dimension for 
various couplings~$J$ in the isotropic Kondo model.}
\end{figure}

In general, new many-particle interactions are generated in
(\ref{floweq}). Our truncation criterion is like in \cite{Kehrein 99} 
the \emph{operator product expansion} (OPE) for vertex operators.
For example in the anticommutator of vertex operators the 
OPE leads to \cite{Delft 98}
\bea \label{OPE}
\lefteqn{\left\{V(\lambda,x), V(-\lambda,y)\right\} = 
\left(1 + i\, \lambda\, (x-y)\, \partial_x \Phi(x) + ...\right)}
\nonumber \\
&&\times
\left(\frac{1}{[1+i (x-y)/ a]^{\lambda^2}} + 
\frac{1}{[1-i (x-y)/ a]^{\lambda^2}} \right) \ .
\eea  
We only keep the leading nonvanishing term of this expansion 
in our calculation. This amounts to neglecting terms with
larger scaling dimensions (more irrelevant terms) and is therefore
reminiscent of an RG approach. Higher order terms can be
successively taken into account in a systematic
expansion. However, the present approximation will already turn out 
to lead to very accurate results.   
Within our approximation, the primary effect of the 
flow (\ref{floweq}) is to generate a new interaction term 
\beq   \label{new_term_1}
H_{\rm new}^{(1)} = S^z \int dx\, f(x)\, \partial_x\Phi(x)
\eeq
in (\ref{HB})
with a function $f(x)$ that depends on the parameters $g_p$. 
(\ref{new_term_1}) is 
identical to the initial longitudinal spin coupling. 
Our strategy 
is to eliminate $H_{\rm new}^{(1)}$ after it is generated infinitesimally
by again performing a unitary transformation 
$U = \exp[i \epsilon S^z \Phi(0)]$ with a suitable 
infinitesimal $\epsilon$.
Similar to the initial value of the scaling dimension that
is determined by the longitudinal Kondo coupling, 
we find a further shift in $\lambda(B)$ due to the elimination 
of $H_{\rm new}^{(1)}$:
one derives the following flow equation 
\beq   \label{lambda_flow}
\partial_B \lambda^2 = 
\frac{8 \pi a \, \lambda^2\, (1 - \lambda^2)}{\Gamma(\lambda^2)}\, 
\sum_p g_p\, g_{-p}\, |p a|^{\lambda^2 - 1} \ .
\eeq
In the sequel we will only look at the strong--coupling phase
($|\Jperp|>-\Jpar$ for small couplings $|\Jperp| \ll 1$) 
since the weak coupling
regime is trivial. In the strong--coupling phase
the flow is always directed towards the \emph{Toulouse line} 
$\lambda = 1$ \cite{Toulouse 69}, which constitutes the \emph{strong--coupling 
fixed point} of our approach (see fig.~\ref{fig:lambda2}).
For $\lambda = 1$ the vertex operators
in (\ref{Hamiltonian_compact}) obey fermionic anticommutation
relations \cite{Delft 98} and the Hamiltonian is quadratic
(equivalent to a noninteracting resonant level model \cite{Toulouse 69}):
The OPE series (\ref{OPE}) terminates after 
the leading term and the flow equations close \emph{exactly} since they
can readily diagonalize a quadratic Hamiltonian. We flow
to a point where our method becomes exact, thereby avoiding the
usual strong--coupling divergence. This is the \emph{key improvement} of
our approach as compared to perturbative scaling, which does not 
make use of the special properties at $\lambda=1$.

The second new interaction that is generated 
is a potential scattering term. Only diagonal terms appear
\begin{displaymath}
H^{(2)}_{\rm new}(B) = \sum_p \omega_{p}(B)\, 
\left(C^\dagger_p\, C^\pdag_p 
- C^\pdag_p\, C^\dagger_p\right)
\end{displaymath}
with the following choice of the generator $\eta^{(2)}_{pq}(B)$
\begin{displaymath}
\eta^{(2)}_{p q} = \frac{1}{2}\, 
\frac{(p+q) \alpha_p \alpha_q g_p g_q}{p - q + 2(\omega_{p} - \omega_{q})} \ .  
\end{displaymath}
$H^{(2)}_{\rm new}(B=\infty)$ is a term in order~$1/L$ 
since $\omega_p(B)=O(1/L)$ and thus represents 
the only "impurity effect" in the Hamiltonian after elimination of the 
Kondo coupling for $B\rightarrow\infty$ in our present 
approximation
\begin{displaymath}
H(B=\infty)=H_0+H^{(2)}_{\rm new}(B=\infty) \ .
\end{displaymath}
Together with (\ref{lambda_flow}) we thus obtain a closed set 
of flow equations
\bea
\partial_B g_p &=& -p^2\, g_p  + 
2 \sum_q \frac{\alpha_q}{\alpha_p}\, \eta^{(2)}_{p q}\, g_q \nonumber \\ 
&& + \frac{1}{2}\, g_p\, \ln\left(\frac{B}{a^2}\right)\, \lambda\, \partial_B \lambda
- 2\, p\, g_p\, \omega_{p}
\label{flow_gp} \\
\partial_B \omega_{p} &=& p\, \alpha_p^2\, g^2_p \ . \nonumber
\eea
Neglecting the momentum dependence induced by $\eta^{(2)}$ and the 
potential scattering terms $\omega_{p}$ (which become important 
only at energies of the order of the Kondo scale), 
we can parametrize the solution of (\ref{flow_gp}) as
$g_p(B)=\tilde{g}(B)\, e^{-p^2 B}$ with the
\emph{running coupling}~$\tilde{g}(B)$. $\tilde{g}(B)$ is in general 
strongly renormalized. 
In the low energy limit the scaling dimension approaches
$\lambda = 1$ and we can then 
deduce the Kondo scale from the effective Toulouse Hamiltonian 
where $T_K \sim \tilde{g}(B=\infty)^2/a$. 
For the symmetric case $\Jpar = \Jperp = J$ 
and a small antiferromagnetic coupling $0<J\ll 1$ one e.g.\ shows
$
T_K \sim \frac{J^{\tau}}{a} \, 
\exp\left(-\frac{2\pi}{J}\right) \ ,
$
where $\tau=(1+\gamma-\ln 2)/3 \approx 0.295$. 
NRG and third order scaling results by Wilson 
\cite{Wilson 75} predict the same leading exponential behaviour 
but a slightly different prefactor with 
$\tau = 1/2$\cite{Cutoff}. 
%
%

\begin{figure}[t]
\begin{center}
\epsfig{file=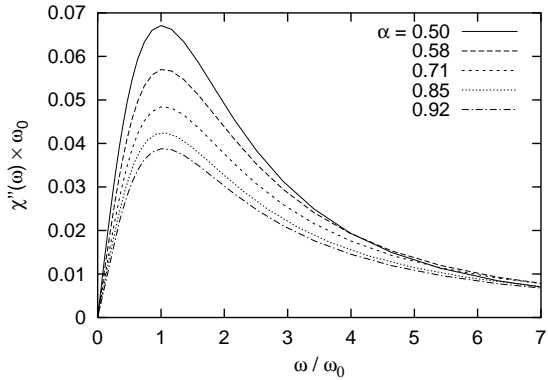,width=0.85\linewidth}
\end{center}
\caption{\label{fig:chi_1}Universal scaling forms of the dynamical impurity susceptibility 
for different dissipation strengths $\alpha = (1 - \Jpar/4\pi)^2$
in the limit of small coupling~$J_\perp$: $\omega_0$ is defined
by the maximum of the curves. One finds $\omega_0\propto T_K$.}
\end{figure}

Our main focus in this 
work is the calculation of the dynamical impurity susceptibility
$\chi(t) = i\, \theta(t)\, \langle[S^z(t),S^z(0)]\rangle$ 
at zero temperature. The key ideas for the calculation of correlation
functions for impurity systems within the flow equation approach have
been developed in~\cite{Kehrein 97a}:
we need to evaluate matrix elements of the 
observable $O = S^z$ with respect to the 
eigenstates of $H(B = \infty)$, \emph{which is simplified by 
the fact that the final Hamiltonian is diagonal.}  
In order to make use of this simplification, we need to 
perform the same series of infinitesimal unitary transformations on $O$ 
as on $H(B)$
\beq  \label{floweq_observable}
\partial_B O(B) = \left[\eta(B),O(B)\right] \ .
\eeq
Usually, keeping the Hamiltonian flow (\ref{floweq}) simple leads to 
a complicated structure of the transformed observable. In the present case, 
however, it 
turns out that the OPE can be used as a unifying truncation 
criterion for \emph{both} $H(B)$ and $O(B)$. Using 
$S^z = [S^+,S^-]/2$ and the ansatz 
\begin{displaymath}
S^+(B) = h(B)\, S^+ + S^z \sum_p d_p(B)\, C^\dagger_p
\end{displaymath}
for the flow of the observable \cite{ansatz}
another closed set of equations is obtained:
\bea  
\partial_B h&=& \frac{1}{2} \sum_p\, p\, \alpha_p \, d_p\, g_p 
\nonumber \\
\partial_B d_p &=& -2\, p\, g_p\, \alpha_p\, h + 
2 \sum_q \eta^{(2)}_{p q}\, d_q \ . \label{floweq_observable_2}
\eea
We have numerically solved this system of equations in combination
with (\ref{flow_gp}).
It is known from the theory of dissipative quantum systems \cite{Kehrein 97a} 
that in the thermodynamic limit the impurity observable~$S^+$ decays 
completely, i.e.~$h(B=\infty)=0$.  
%
%
With $S^z(B=\infty)$ completely expressed in terms 
of band operators, it is then trivial to calculate the 
dynamical susceptibility: since $H^{(2)}_{\rm new}(B=\infty)$
is of order~$1/L$, it does not contribute to the dynamics and
we need only study the time evolution of $S^z(B=\infty)$ under $H_0$.
The imaginary part of the Fourier-transform of $\chi(t)$ follows
easily 
\bea
\chi''(\omega) &=& \frac{\pi}{16} \sum_{p,q > 0} 
d^2_p(B=\infty)\, d^2_q(B=\infty) \nonumber \\
&&\qquad\times\left(\delta(\omega - \epsilon_p - \epsilon_q) - 
\delta(\omega + \epsilon_p + \epsilon_q)\right) \ . \nonumber
\eea
Results are shown in fig.~\ref{fig:chi_1}: The curves display 
a broad maximum at an energy of the order of the 
Kondo scale and a power law decay 
$\chi''(\omega) \sim \omega^{-(3 - 2 \alpha)}$
at high frequencies, consistent with results obtained 
for the spin boson model \cite{Costi 99}. 
At low frequencies one finds $\chi''(\omega) \sim \omega$,  
where the corresponding slope scales as  
$T_K^{-2}$ with a prefactor of order one \cite{Slope}.
The normalization condition 
$\int_0^\infty d\omega\, \chi''(\omega) = \pi/4$
is fulfilled due to the sum rule 
$h^2(B) + \sum_p d^2_p(B)/4 = 1$ that holds exactly \cite{Wegner 99}.

The real part $\chi'(\omega)$ can be obtained by a Kramers-Kronig 
transformation. 
In particular, the \emph{static} susceptibility in response 
to a local field is given by 
$\chi_0 = \chi'(\omega=0)$
and therefore contains spectral information on all energy scales.
In the limit of small spin flip coupling 
$\Jperp$ we have numerically verified the power law behaviour 
$\chi_0 \sim \Jperp^{1/(\alpha - 1)}$ known from 
the ohmic spin boson model with the 
dissipation parameter $\alpha = (1 - \Jpar/4\pi)^2$ \cite{Costi 99}.
%
%
%
%

Summing up, we have applied Wegner's 
flow equations to the anisotropic Kondo problem. 
The Hamiltonian is written in terms of vertex operators 
with scaling dimensions that flow to the Toulouse point.
In contrast to the perturbative scaling approach,  
our flow equations close exactly at the  
Toulouse point with finite couplings and a nontrivial 
strong--coupling behaviour. Similar to the analysis of the
sine--Gordon model \cite{Kehrein 99}
no strong--coupling divergence of the coupling constants is encountered.
We are thus able to give an analytic description of 
the crossover from weak to strong coupling in a systematic
expansion that can be improved by taking higher orders 
in the OPE into account. The stability of the strong--coupling
fixed point cannot be endangered in this expansion.

In our solution of the flow equations we have focussed on
the equilibrium dynamics of the impurity at zero temperature. 
As an example we have calculated the universal scaling forms of 
the local dynamical susceptibility
$\chi(\omega)$ and found good agreement 
with results known from the spin-boson model.
This exemplifies the usefulness of our approach
for the calculation of correlation functions in strong--coupling 
problems, and should be of interest in other 
models as well. Notice that we have not used the integrable structure
of the Kondo model in our method. 
Future work will concentrate on the effect of finite temperature, and
on the calculation of static impurity 
contributions where the newly generated potential scattering 
terms give the leading contribution (in contrast to the 
dynamical susceptibility where these terms play no role) and our 
present approximation must be improved. 

The authors acknowledge many valuable discussions with T.A.~Costi,
D.~Fisher, A.~Mielke and D.~Vollhardt.
This work was supported by the Deutsche Forschungsgemeinschaft (DFG), by
the SFB~484 of the DFG and by the National Science Foundation (NSF) under 
grants DMR~9630064, DMR~9976621 and DMR~9981283.


\begin{thebibliography}{99}
\bibitem{Kleinfactors} The Klein factors can be absorbed into 
redefined spin operators and are of no concern in the sequel.
\bibitem{Cutoff}Possible reasons for this discrepancy are either the 
neglected higher order OPE terms, or the different cutoff schemes 
in the fermionic and bosonized models.
\bibitem{ansatz}Again more irrelevant terms are neglected in this
ansatz, which becomes exact on the Toulouse line.
\bibitem{Slope}This prefactor of order one is not obtained correctly in the 
present approach. However, this discrepancy occurs on an energy
scale $\ll \omega_0$ and cannot be resolved in fig.~2. 
The deviation is due to our approximations in this work and will
be improved in a subsequent publication.
\end{thebibliography}
\end{document}